\documentclass[prl,twocolumn,superscriptaddress]{revtex4}
\usepackage{color}
\usepackage{graphicx}
\usepackage{graphics}
\begin{document}

\title{ Structural stability of the $B_{80}$  fullerene against defect formation}

\author{Kuo Bao}
\affiliation{ State Key Lab of Superhard Materials, Jilin
University, Changchun 130012, China \\} \affiliation{Department of
Physics, Universit\"{a}t Basel, Klingelbergstr. 82, 4056 Basel,
Switzerland \\}
\author{Stefan Goedecker}
\affiliation{Department of Physics, Universit\"{a}t Basel,
Klingelbergstr. 82, 4056 Basel, Switzerland \\}
\author{ Luigi Genovese }
\affiliation{CEA, INAC, SP2M, Laboratoire de simulation atomistique (L\_Sim), F-38054 Grenoble, France \\}
\affiliation{European Synchrotron Radiation Facility, BP220, F-38043 Grenoble Cedex, France \\}
\author{Alexey Neelov}
\affiliation{Department of Physics, Universit\"{a}t Basel,
Klingelbergstr. 82, 4056 Basel, Switzerland \\}
\author{Alireza Ghasemi}
\affiliation{Department of Physics, Universit\"{a}t Basel,
Klingelbergstr. 82, 4056 Basel, Switzerland \\}
\author{ Thierry Deutsch}
\affiliation{CEA, INAC, SP2M, Laboratoire de simulation atomistique (L\_Sim), F-38054 Grenoble, France \\}

\begin{abstract}
Using a systematic search algorithm we identify several types of 
point defects in the boron fullerene with 80 atoms. 
All these point defect leave the cage structure intact.
In addition the cage structure is also very stable 
with respect to elastic deformations and
addition or removal of atoms.
\end{abstract}

\pacs{PACS numbers: ?? }

\maketitle

Since boron is a neighbor of carbon in the periodic table, they
share a lot of similarities. The $sp^2$ hybridization of the valence
electrons, the electron deficiency, large coordination numbers and
short covalent radius allow boron to form strong directed chemical
bonds and give rise to rich variety of possible nanostructures.
Boron nanostructures share some common basic units following the
so-called "Aufbau principle"\cite{ihsan}. Various kinds of boron
nanostructures, such as boron nanotubes\cite{ntubes,tsinghua}, boron
fullerenes \cite{nevill,india} and boron sheets\cite{tsinghua} have
been studied both experimentally and theoretically. The $B_{80}$
fullerene has recently been identified in density functional
calculations as an energetically very favorable
structure\cite{polish}. Like the carbon fullerenes boron based
fullerenes might thus play an important role\cite{american} if one
succeeds in synthesizing them. In contrast to other fullerenes where 
defects have been studied in detail, the defects in $B_{80}$ have 
not yet been studied. Since materials are never perfect in reality, it 
is important to have information about the structure and 
energetics of their defects. in this article we study the defects 
of $B_{80}$. Whereas in previous studies of defects in various 
materials educated guesses were made to identify defects, we 
use a systematic search algorithm~\cite{minhop} to find them. 
As a consequence we find not only one type of defect but several 
of them. All the defects we found leave the cage structure intact.
This stability should make their
synthesis easier. If in an initial stage of the synthesis an
imperfect structure is generated, it forms nevertheless a cage
structure and it can later anneal into the perfect cage structure.

\begin{figure}[h]             
\begin{center}
\setlength{\unitlength}{1cm}
\includegraphics[width=5cm]{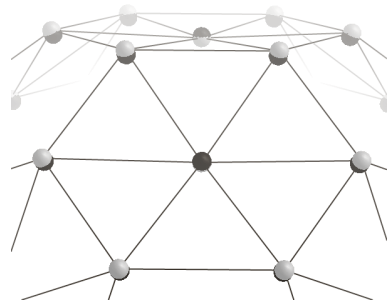}   
\caption{ \label{both} A section of the $B_{80}$ fullerene that illustrates the differences
in the ground state geometries of the LDA ground state (grey spheres) and the
PBE ground state (black spheres). In the LDA geometry the hexagon centered atoms are exactly
in the plane of the hexagon, in the PBE geometry they move slightly inside.}
\end{center}
\end{figure}

In the first publication on the $B_{80}$ fullerene a perfect
icosahedral structure was proposed\cite{polish}. In a subsequent
publication\cite{american} a distorted icosahedron was presented as
the ground state if the PBE\cite{pbe} density functional is used.
Our calculations confirm that the distorted cage is the ground state
within PBE. However the distortion is marginal and atoms are
displaced by less than 0.035 Bohr with respect to the ideal
structure as shown in Fig~\ref{both}. The energy differences are
also small and as a matter of  fact a LDA functional~\cite{gth}
favors the perfect icosahedral structure by about the same amount of
energy (Table~\ref{energies}).
Because the deviations from
the icosahedral structure are so small we will call in the following
both the perfect and the slightly distorted icosahedral states just
icosahedral.\

In order to search in a systematic way for defect structures we used
the minima hopping algorithm~\cite{minhop} together with the BigDFT
electronic structure program~\cite{bigdft} from the ABINIT
package~\cite{abinit}. The BigDFT electronic structure program uses
a wavelet basis set. This is a systematic basis set and the number
of basis functions was chosen such that energies are accurate to
within better than 1 mHa.  Pseudopotentials were used to eliminate
the core electrons~\cite{hgh}. Minima hopping is an algorithm to
find the global minimum as well as other low energy structures. It
is entirely unbiased and applicable to any system. After exploring
some 20 configurations with the minima hopping algorithm we could
not find any configurations where the cage structure was destroyed.
The configurations we found in this way were combinations of several
of the basic defect structures that we will discuss below. Combining
these basic defects can give rise to a large number of defective
cage structures all of which are very close in energy to the ground
state.

\begin{figure}[h]             
\setlength{\unitlength}{1mm}
\begin{picture}(50,180)
\put(5,140){\includegraphics[width=5cm]{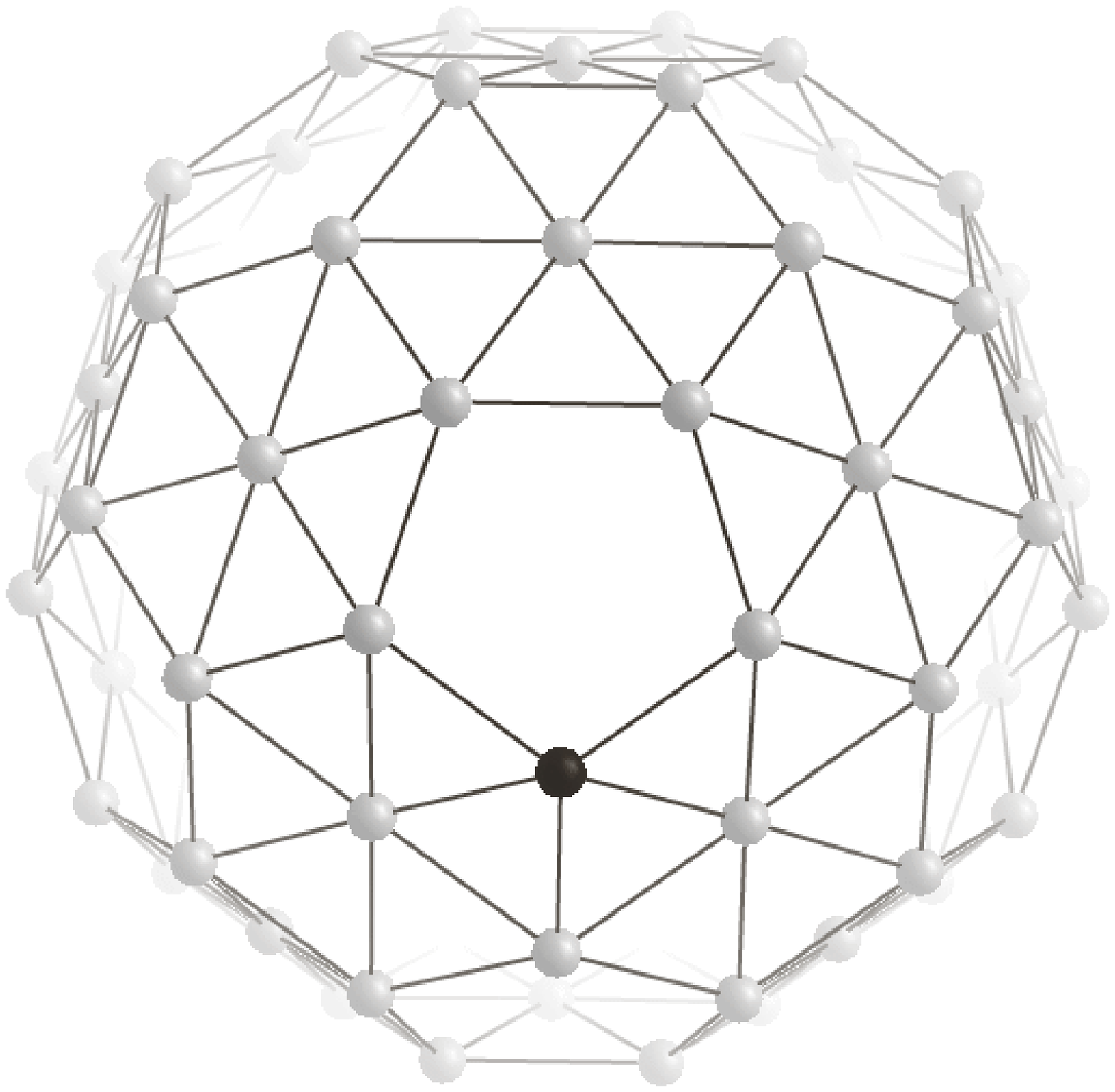}}   
\put(5, 95){\includegraphics[width=5cm]{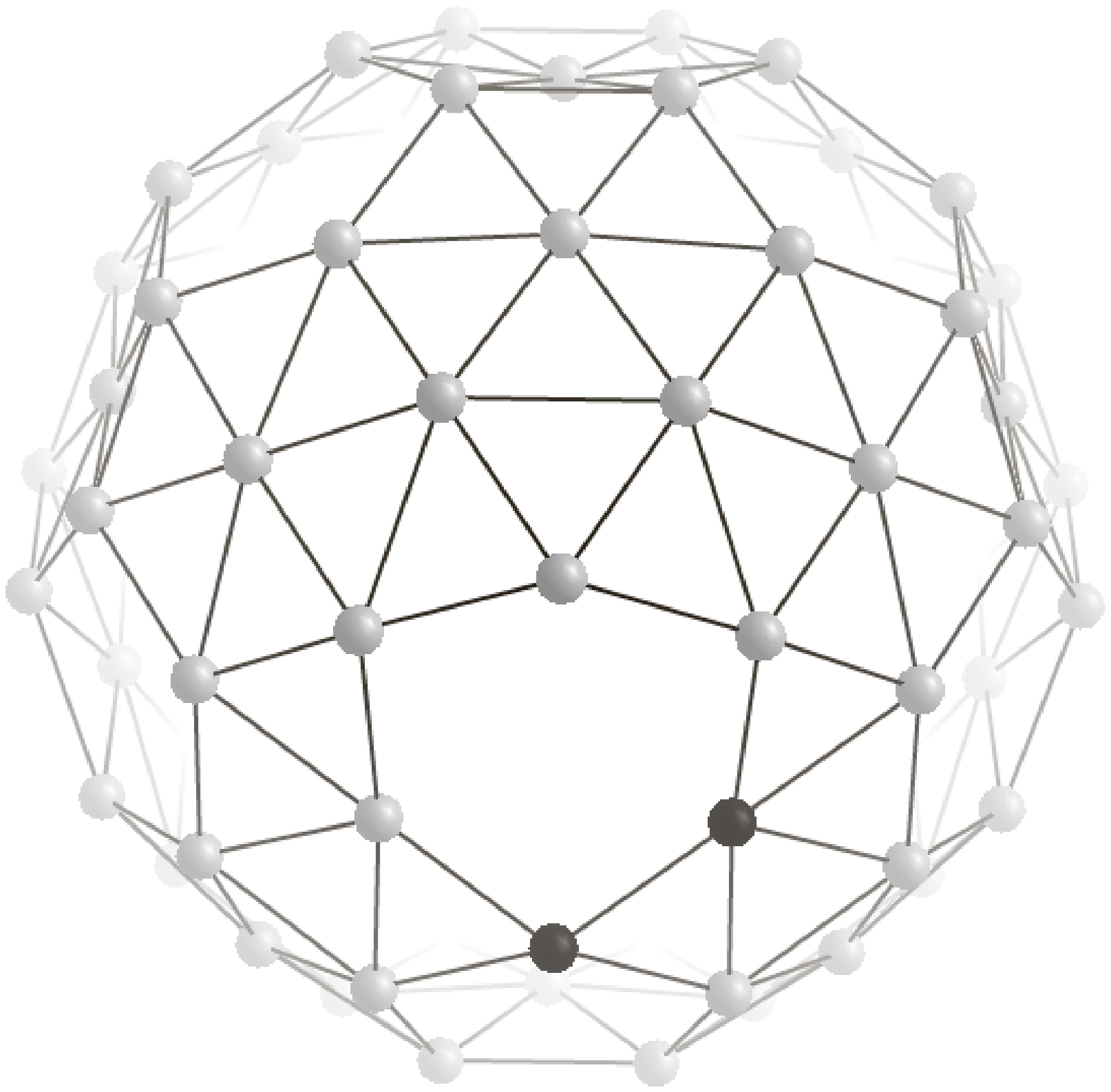}}   
\put(5, 50){\includegraphics[width=5cm]{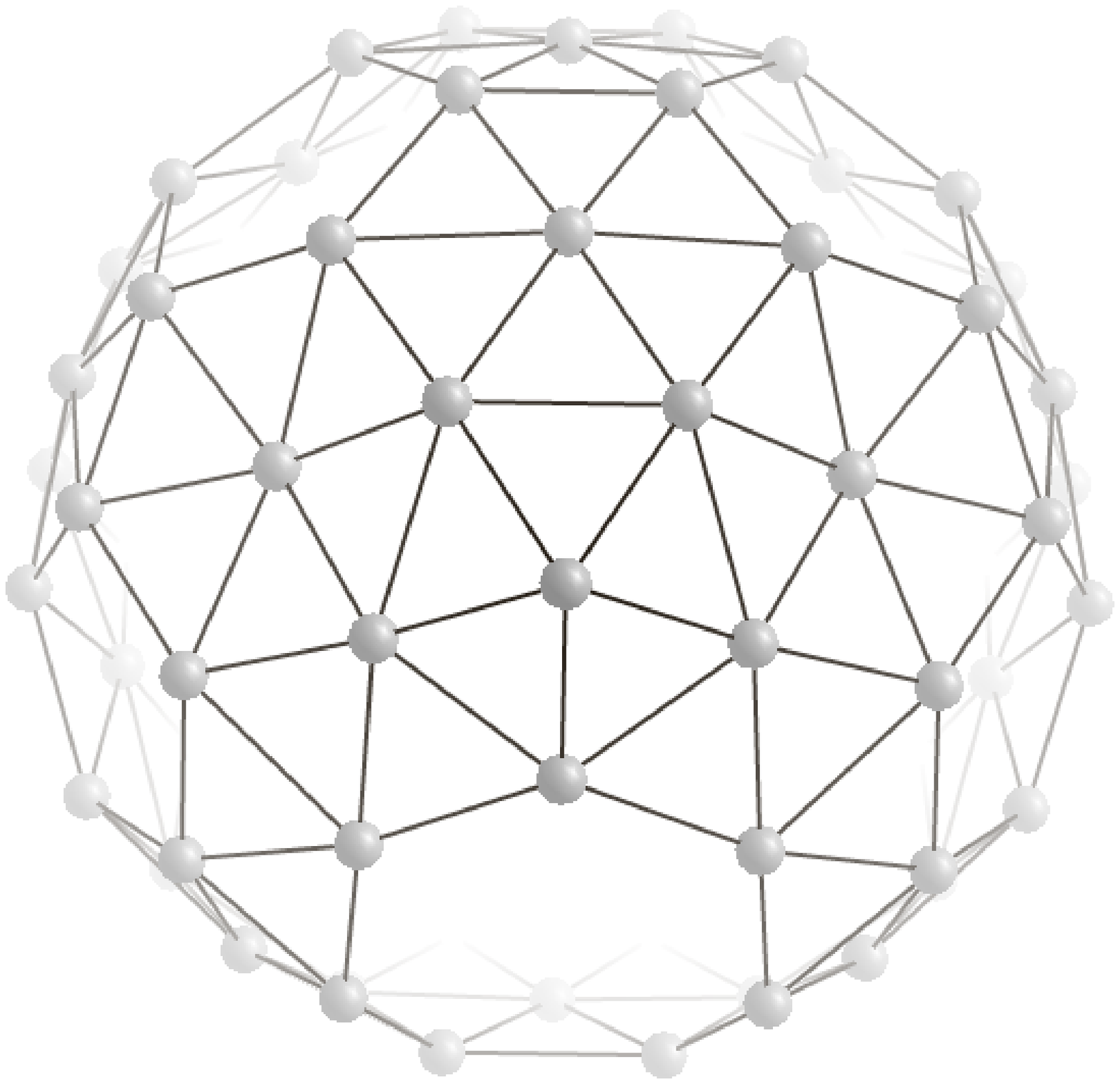}}   
\put(5,  5){\includegraphics[width=5cm]{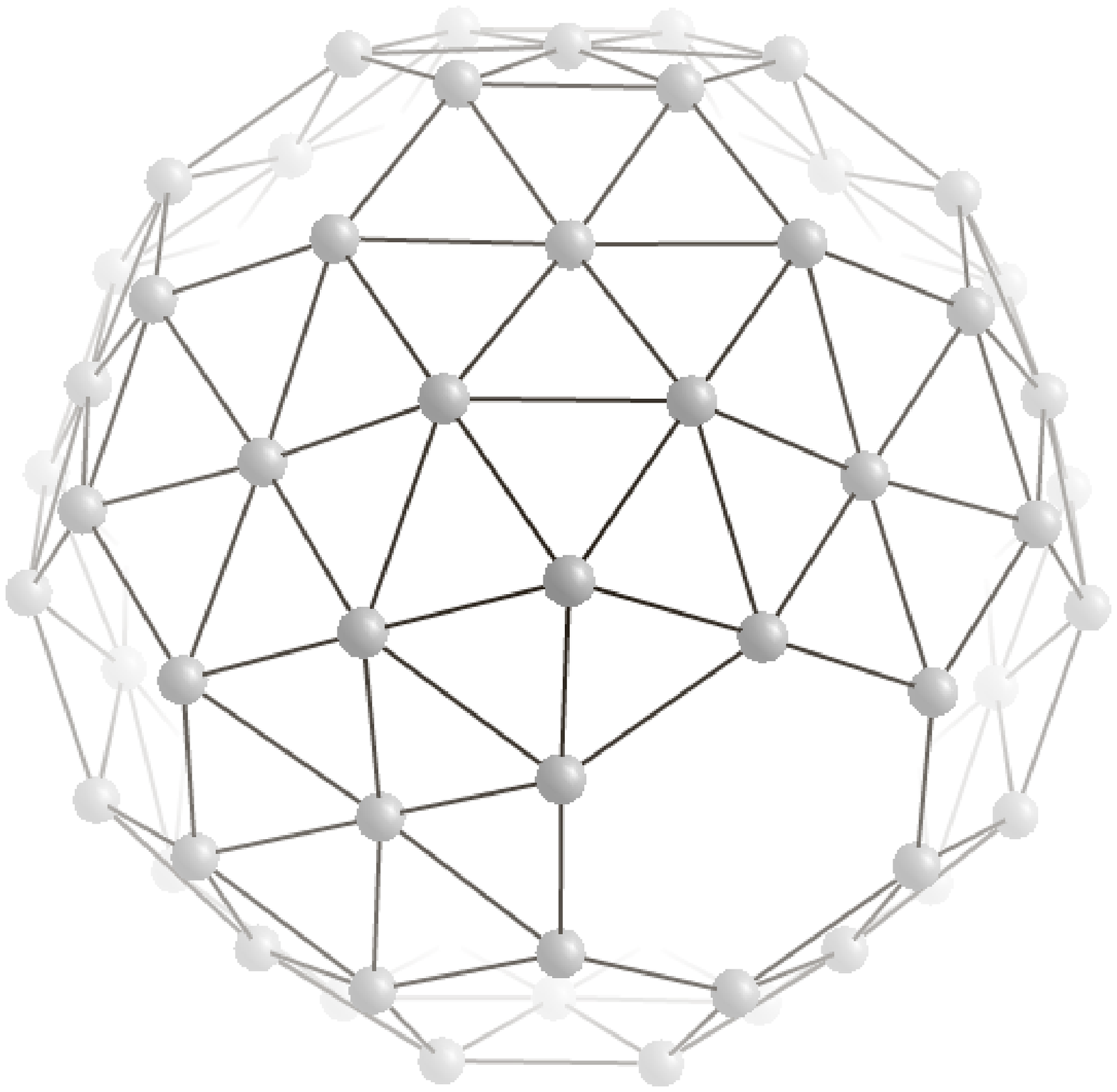}}   
\put(2,180){\hbox{\large A}} \put(2,125){\hbox{B}}
\put(2,90){\hbox{C}} \put(2,45){\hbox{D}}
\put(30,158){\linethickness{1pt}{\vector(0,1){4}}}
\put(30,106){\linethickness{1pt}{\vector(0,1){4}}}
\put(36,110){\linethickness{1pt}{\vector(-1,1){4}}}
\end{picture}
 \caption{ \label{defects} Illustration of the hexagon and heptagon
defects. To get from the icosahedral ground state configuration (top
panel A) to the hexagon defect structure (panel below) one has to
shift the black atom in the uppermost panel upward as indicated by the arrow. The hexagon
defect in the second lowest panel B can then be shifted to the
positions in the second lowest and lowest panel C and D by moving upwards or
upwards to the left either of the two black atoms in the second
uppermost panel as indicated by the two arrows. In panel B and D the defect is
a hexagon, in panel C a heptagon. The defects in panels B,C and D will be denoted in
Table~\ref{energies} as type 1,2 and 3.}
\end{figure}

A $B_{80}$ fullerene can be obtained from the $C_{60}$ fullerene structure
 by adding 20 atoms in the center of the 20 hexagons of the  $C_{60}$ fullerene structure.
The $B_{80}$ fullerene exhibits however a larger variety of defect
structures than the  $C_{60}$ fullerene. There are defects where a
pentagon is destroyed and a distorted hexagon or heptagon is created. We will in
the following call this new class of defects hexagon/heptagon defects. Once a
defect has been created it can be shifted to different places on
the surface of the cage. This is shown in Fig.~\ref{defects}. The
barriers for the displacements of the hexagons are low. For the
process leading from the perfect structure (top structure in
Fig.~\ref{defects}) to the first defect structure (below perfect
structure in Fig.~\ref{defects}) the barrier height is .073 Ha (2.0 eV) in LDA
and .067 Ha (1.8 eV) in PBE. Displacements of defects are thus expected to
give rise to some plastic deformation behavior of the $B_{80}$
fullerene.

\begin{figure}
\begin{center}
\setlength{\unitlength}{1cm}
\includegraphics[width=5cm]{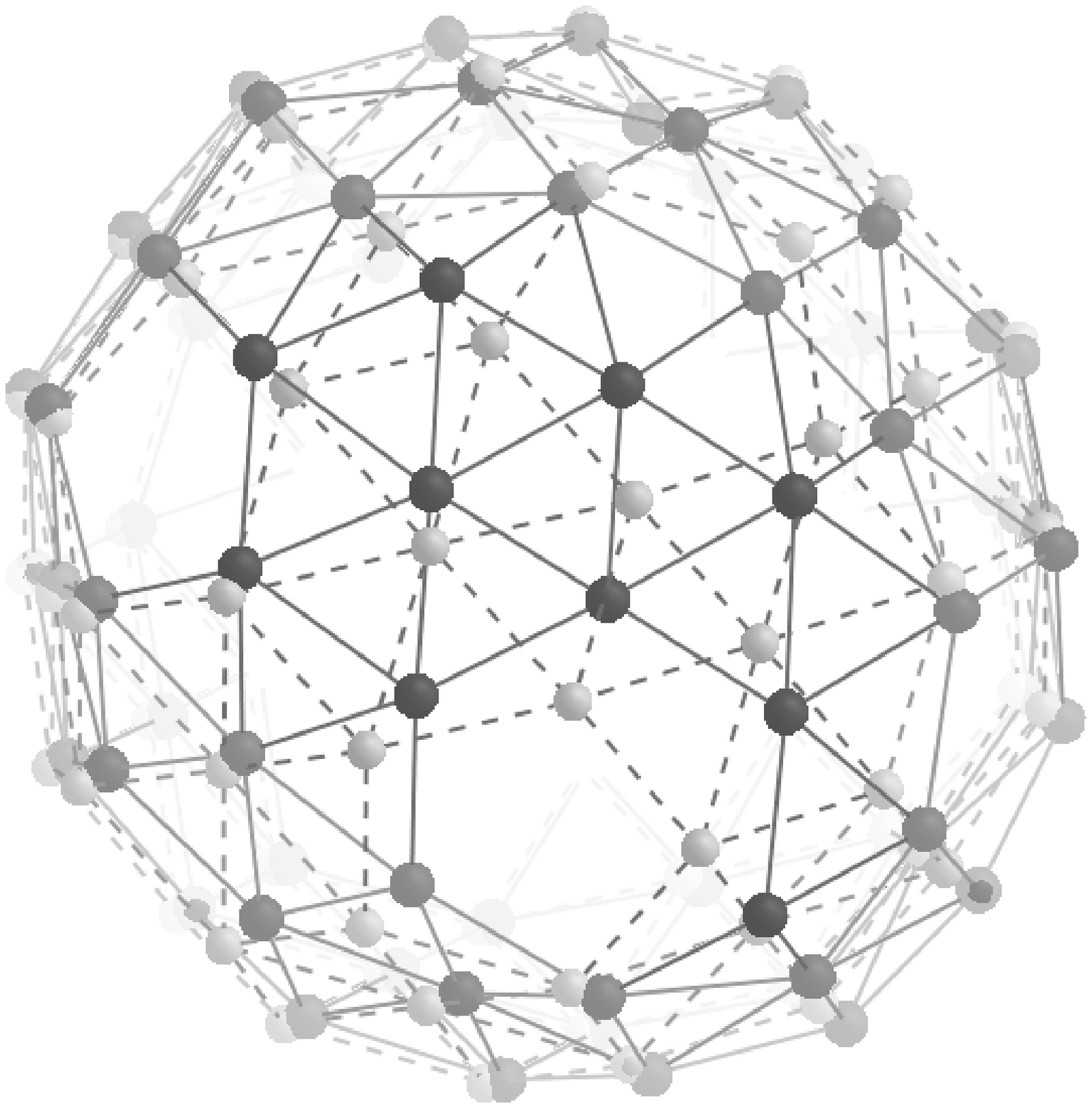}   
\caption{ \label{crashing} The picture shows the most distorted part of
a type 4 defect in $B_{80}$. The lighter atoms and dotted bonds
show the positions of atoms in perfect structure and the dark atoms
with solid bonds shows the B$_{80}$ with the defect.}
\end{center}
\end{figure}

Hexagon defects can also be created by shifting a group of atoms as
shown in Fig.~\ref{crashing}. In contrast to the previous hexagon
defect shown in panel D of Fig.~\ref{defects}, where the hexagon is
neighbor to a pentagon, the hexagon is isolated from the pentagons
in this case. The fullerene cage is also relatively strongly
distorted for this defect.  we will call this defect type 4 defect.

\begin{figure}[ht]             
\setlength{\unitlength}{1mm}
\begin{picture}(70, 45)
\put(-5,5){\includegraphics[width=3cm]{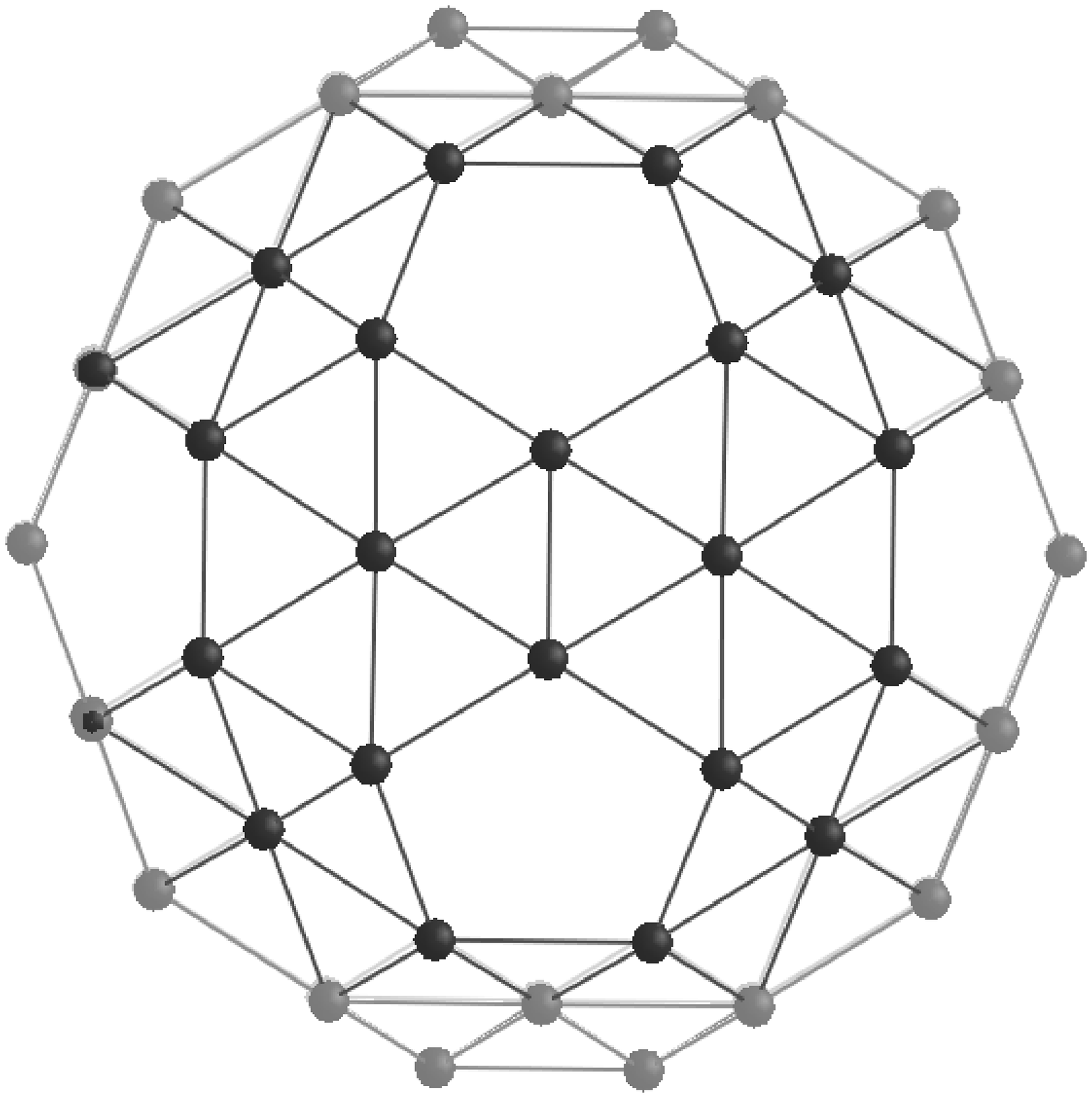}}   
\put(40,5){\includegraphics[width=3cm]{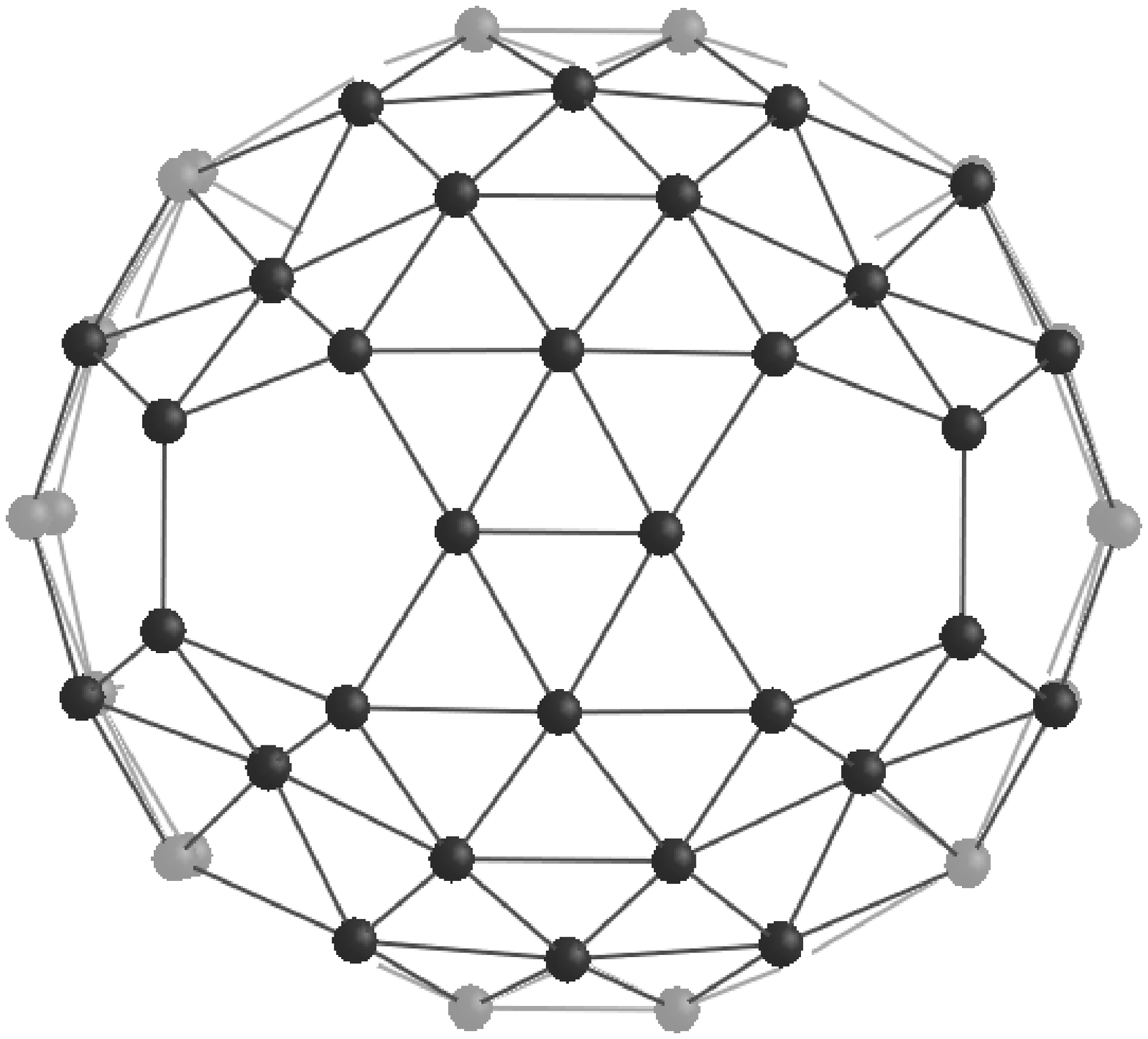}}        
\put(34,25){\linethickness{1pt}{\vector(1,0){6}}}
\end{picture}
\caption{ \label{stone} A perfect B$_{80}$ (on the left) and a
Stone-Wales type defect in $B_{80}$.}
\end{figure}

It is not too surprising that the $B_{80}$ fullerene
inherits also the Stone-Wales type of defect~\cite{stone}. If one
adds to a $C_{60}$ with a Stone-Wales defect again 20 atoms into the
hexagons one obtains the $B_{80}$  with a Stone-Wales defect shown
in Fig.~\ref{stone}. The number of pentagons is 12 as in the perfect
structure, but two pentagons became now neighbors. More pathways exist 
however in $B_{80}$ than in $C_{60}$ to create a Stone-Wales defect.
One can for instance transform the defect in Fig.~\ref{crashing} into a 
Stone-Wales defect by moving one atom which is part of the hexagon and 
closest to the next pentagon towards the  center of the hexagon. 

\begin{table}[h]
\begin{tabular}{|c|c|c|}
 \hline                   &   LDA   & PBE            \\ \hline \hline
  perfect Icosahedron     & 0.      & 0.     \\ \hline
  distorted Icosahedron   & .020    & -.020  \\ \hline
  defect type 1           & .053 & .027     \\ \hline
  defect type 2           & .062 & .035     \\ \hline
  defect type 3           & .027 & .066     \\ \hline
  defect type 4           & .061 & .040     \\ \hline
 Stone Wales type defect  & .050 & .026     \\ \hline
\end{tabular}
\caption{The energies of various configurations with respect to the
highest symmetry icosahedral fullerene as calculated with the
LDA~\cite{gth} and PBE~\cite{pbe} density functionals in unit of Ha.
All the energies are the energies of relaxed structures with the
exception of perfect icosahedron in PBE and the distorted
icosahedron in LDA. \label{energies}}
\end{table}

\begin{figure}[ht]             
\begin{center}
\setlength{\unitlength}{1cm}
\includegraphics[width=5cm]{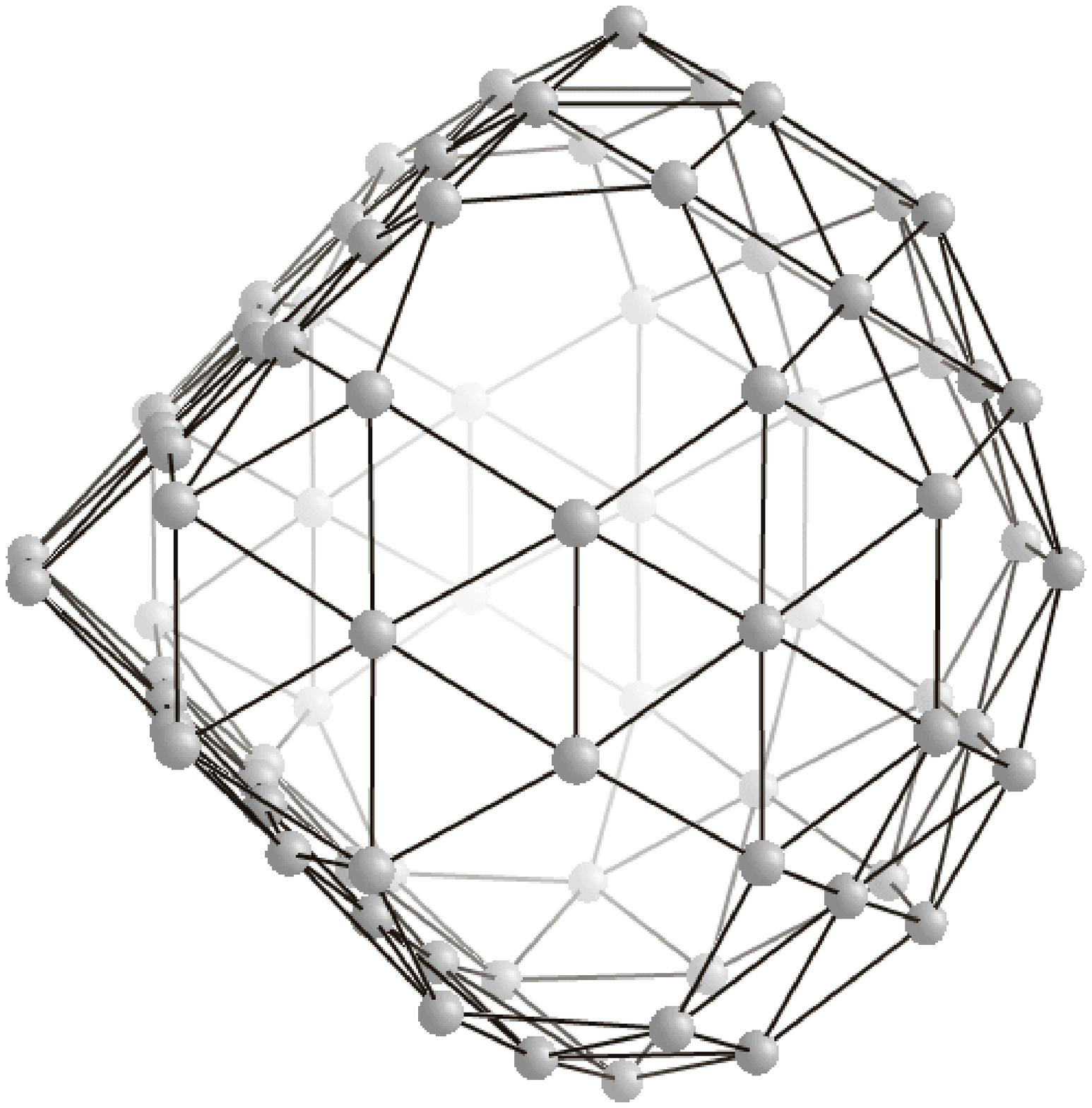}   
\caption{ \label{corner} A $B_{80}$ fullerene that was deformed such
that it has a sharp edge and two flat surfaces. This strongly
deformed structure relaxes back to the icosahedral structure. }
\end{center}
\end{figure}

In addition to being stable against defect formation, the $B_{80}$
fullerene is also highly stable against elastic deformations. We
enforced for instance several strong deformations onto the cage such
as a sharp edge together with flat regions shown in
Fig.~\ref{corner}. Such a structure could serve as a nucleation site
for a more compact cluster structure. We found however that the
fullerene always relaxed back to the open icosahedral structure
during a geometry optimization.

The $B_{80}$ fullerene is also stable against the removal or
addition of a single atom. If one removes a boron atom from the
center of a hexagon to  obtain $B_{79}$ or if one adds a boron atom
in the center of a pentagon to obtain $B_{81}$ one obtains
structures whose cohesive energy per atom differs by less than .5
mHa from the values of the icosahedral $B_{80}$. Consecutive
hexagons can be filled until one obtains the $B_{92}$ fullerene. The
cohesive energies per atom changes hardly when going from $B_{80}$
to $B_{92}$. The cohesive energy for $B_{92}$ that we calculated is
only 1 mHa less than the one for $B_{80}$. Our number is exactly in
between the two numbers given in reference~\cite{polish}.

In summary, we have shown that several different types of point defects exist
in the $B_{80}$ fullerene and that the cage like structure is stable against
all these point defects as well as against strong deformations and changes in the
number of atoms.  Since other boron nanostructures are built up according
to the same basic construction principles, it is to be expected that the same or
very similar defects can be found in other boron nanostructures.

We thank Mark Pederson and Tunna Baruah for providing us with the $B_{80}$ ground state structure.
Financial support was provided by the Swiss National
Science Foundation and the CSC (China Scholarship Council).
The calculations were done at the Swiss National Supercomputing Center CSCS.
For visualizations the v\_sim software was used (http://inac.cea.fr/L\_Sim/V\_Sim).

\end{document}